\RequirePackage{ifpdf}
\ifpdf 
\documentclass[pdftex]{sigma}
\else
\documentclass{sigma}
\fi

\usepackage{epsfig}

\def\be{\begin{equation}}
\def\ee{\end{equation}}

\def\bea{\begin{eqnarray}}
\def\eea{\end{eqnarray}}
\renewcommand\L{\mathcal{L}}
\def\G{\mathcal{G}}

\def\arrowsite{\begin{picture}(45,5)(-2,-2)
\put(0,0){\circle*{5}} \put(2.5,0){\vector(1,0){10}}
\put(12.5,0){\line(1,0){15}} \put(37.5,0){\vector(-1,0){10}}
\put(40,0){\circle*{5}}
\end{picture}}

\def\arrowroot{\begin{picture}(45,5)(-2,-2)
\put(0,0){\circle*{5}} \put(2.5,0){\vector(1,0){10}}
\put(15,0){\circle{5}} \put(25,0){\circle{5}}
\put(37.5,0){\vector(-1,0){10}} \put(40,0){\circle*{5}}
\end{picture}}

\begin{document}

\allowdisplaybreaks

\renewcommand{\PaperNumber}{001}

\FirstPageHeading

\renewcommand{\thefootnote}{$\star$}

\ShortArticleName{Non-Local Finite-Size Ef\/fects in the Dimer
Model}

\ArticleName{Non-Local Finite-Size Ef\/fects in the Dimer
Model\footnote{This paper is a contribution to the Proceedings of
the O'Raifeartaigh Symposium on Non-Perturbative and Symmetry
Methods in Field Theory
 (June 22--24, 2006, Budapest, Hungary).
The full collection is available at
\href{http://www.emis.de/journals/SIGMA/LOR2006.html}{http://www.emis.de/journals/SIGMA/LOR2006.html}}}

\Author{Nickolay Sh. IZMAILIAN~$^{\dag^1\dag^2\dag^3}$,
 Vyatcheslav B. PRIEZZHEV~$^{\dag^4}$ and Philippe RUELLE~$^{\dag^5}$}

\AuthorNameForHeading{N.Sh. Izmailian, V.B. Priezzhev and P. Ruelle}

\Address{$^{\dag^1}$~Institute of Physics, Academia Sinica,
Nankang, Taipei 11529, Taiwan}
\EmailDD{\href{mailto:izmailan@phys.sinica.edu.tw}{izmailan@phys.sinica.edu.tw}}
\Address{$^{\dag^2}$~Yerevan Physics Institute, Alikhanian
Brothers 2, 375036 Yerevan, Armenia} 

\Address{$^{\dag^3}$~National Center of Theoretical Sciences at Taipei, Physics Division,\\
$\phantom{^{\dag^3}}$~National Taiwan University, Taipei 10617,
Taiwan} \Address{$^{\dag^4}$~Bogolyubov Laboratory of
Theoretical Physics, Joint Institute for Nuclear Research,\\
$\phantom{^{\dag^4}}$~141980 Dubna, Russia}
\EmailDD{\href{mailto:priezzvb@theor.jinr.ru}{priezzvb@theor.jinr.ru}}
\Address{$^{\dag^5}$~Institut de Physique Th\'eorique, Universit\'e catholique de Louvain, \\
$\phantom{^{\dag^5}}$~1348 Louvain-La-Neuve, Belgium}
\EmailDD{\href{mailto:ruelle@fyma.ucl.ac.be}{ruelle@fyma.ucl.ac.be}}

\ArticleDates{Received September 29, 2006, in f\/inal form
December 12, 2006; Published online January 04, 2007}

\Abstract{We study the f\/inite-size corrections of the dimer
model on $\infty \times N$ square lattice with two dif\/ferent
boundary conditions: free and periodic. We f\/ind that the
f\/inite-size corrections depend in a crucial way on the parity of
$N$, and show that, because of certain non-local features present
in the model, a change of parity of $N$ induces a change of
boundary condition. Taking a careful account of this, these
unusual f\/inite-size behaviours can be fully explained in the
framework of the $c=-2$ logarithmic conformal f\/ield theory.}


\Classification{82B20; 81T40} 

\section{Introduction}

The dimer model is extremely simple to def\/ine. We take a
rectangular grid $\L$ with $M$ rows and $N$ columns, and consider
all arrangements of dimers (dominoes) so that all sites of $\L$
are covered by exactly one dimer. An example of such an
arrangement for a $9 \times 18$ grid is shown in Fig.~1a. In case
both $M$ and $N$ are odd, one site has to be removed from the
grid; this site is conventionally to be one of the four corners.

One can be more general and introduce monomers (vacancies), namely
f\/ixed sites which cannot be covered by dimers. A dimer
arrangement is possible in presence of $m$ monomers provided the
numbers of sites which must be covered by dimers and belonging to
the even sublattice and to the odd sublattice (sum of coordinates
even or odd resp.) are equal. In particular, $MN-m$ must be even.
A dimer conf\/iguration on a $9 \times 18$ grid with four monomers
is shown in Fig.~1b. The case $m=0$ (or $m=1$ if $MN$ is odd) is
referred to as the close-packed limit of the dimer model.
Corresponding to these arrangements of monomers and dimers, we
introduce the partition function
 \be Z(x,y|z_1,\ldots,z_m) =
\sum_{\rm coverings} x^{n_h}  y^{n_v}. \label{pf} \ee 
It counts
the number of dimer coverings in presence of $m$ monomers located
at positions $z_1, \ldots,z_m$, in the bulk or on boundaries, with
respective weights $x$ and $y$ assigned to horizontal and vertical
dimers. As the number $n_h + n_v$ of dimers is f\/ixed, the
partition function essentially depends on $x$, $y$ through the
ratio $x/y$ only

\begin{figure}[t]
\includegraphics[width = 7.1cm]{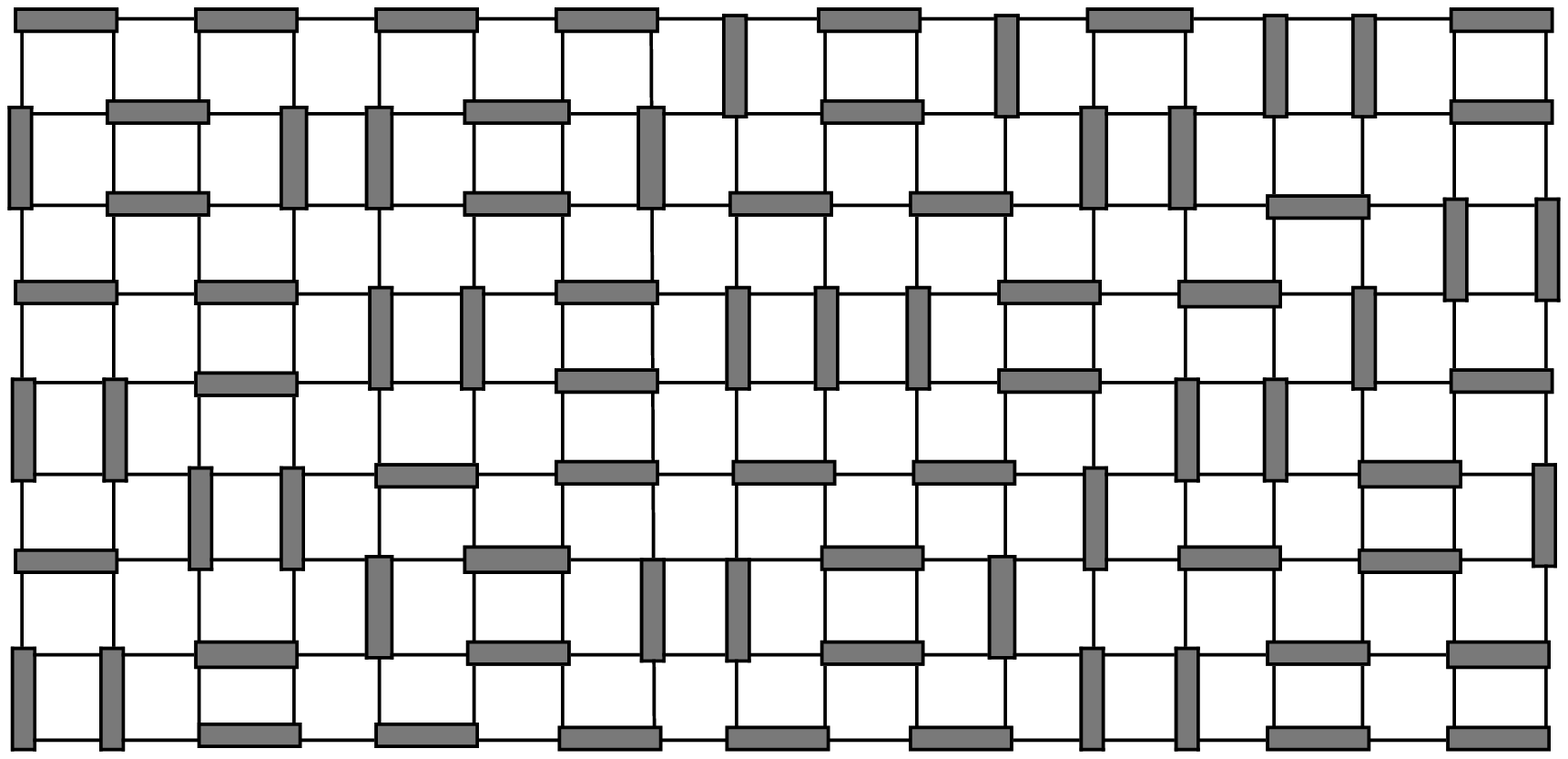}
\hfill
\includegraphics[width = 7.1cm]{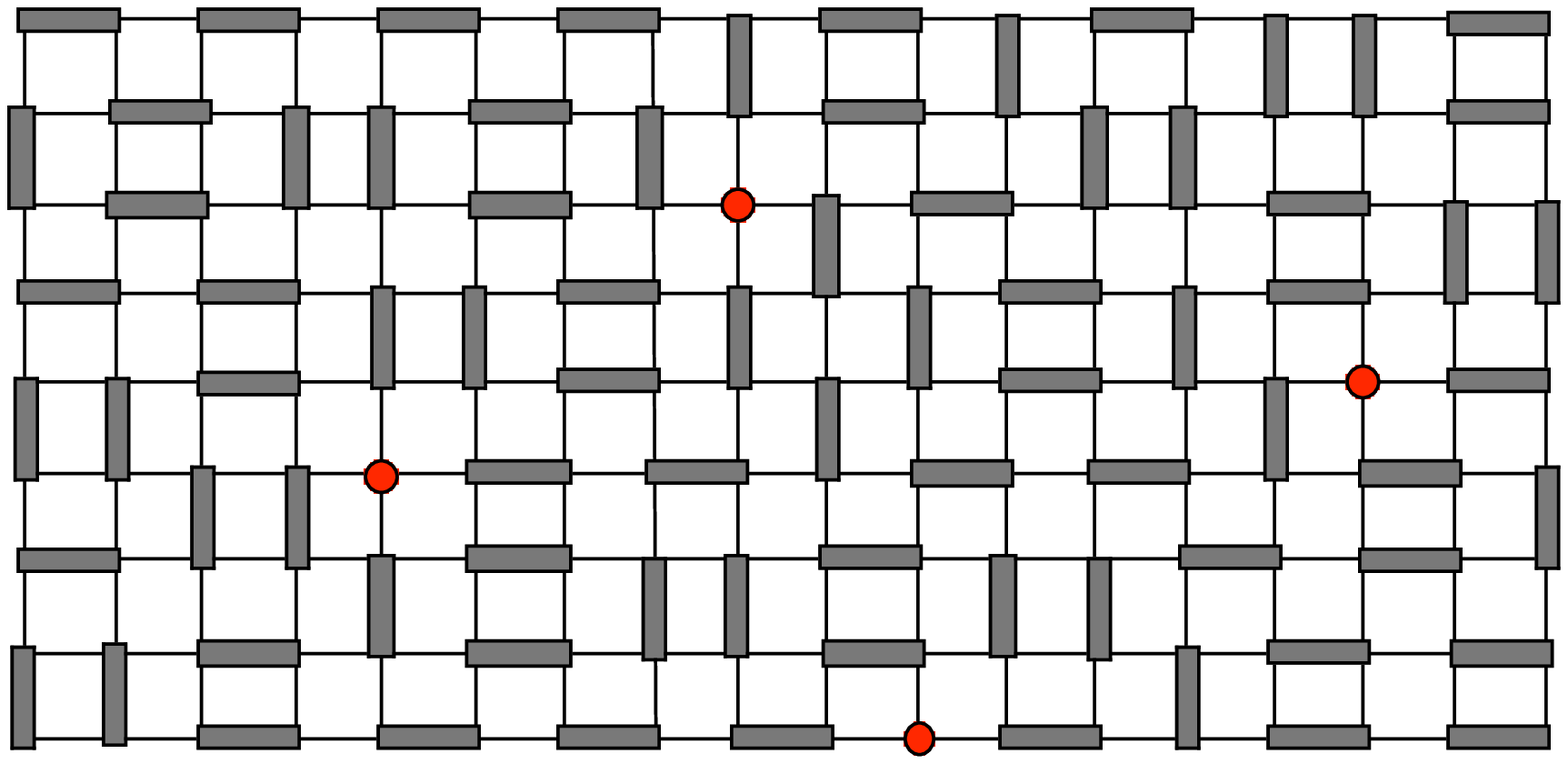}
\caption{(a) The left f\/igure shows one among some $4.653 \times
10^{18}$ close-packed dimer coverings on a $9 \times 18$ grid. (b)
The right f\/igure shows a~dimer conf\/iguration on the same grid
but with four monomers, represented by red dots. The insertion of
the four monomers has reduced the number of conf\/igurations by a
factor roughly equal to 57.5, down to $8.097 \times 10^{16}$.}
\end{figure}

There are obvious generalizations (other types of lattices,
coverings by other lattice animals), but the basic questions
remain the same. Depending on the values of the parameters (here,
$x/y$), can the system become critical and does it exhibit phase
transitions? What is the probability that certain chosen bonds be
covered by dimers? How do monomers af\/fect the number of dimer
coverings? More importantly for our purpose here, what are the
f\/inite-size correction terms in the partition function, and, in
case the system is critical, is its scaling regime conformally
invariant?

The dimer model has been originally introduced to model physical
adsorption of diatomic molecules on crystal surfaces \cite{foru}.
The f\/irst studies of the dimer model were conducted in the early
sixties, with pioneering works by Kasteleyn \cite{kast1,kast2},
Fisher \cite{fish}, and Fisher and Temperley~\cite{fite}. Soon
after that, correlations between dimers and monomers on the square
lattice have been studied in \cite{fist,hart}, and have been
revisited more recently for the triangular lattice in
\cite{fend,baeh}. The ef\/fects caused by the insertion of
monomers have also been reconsidered recently \cite{tzwu,kong,wu}.
Finite-size ef\/fects also have a long history, starting in
\cite{fish,ferd}, with many subsequent works
\cite{coywu,bhana,brpr,luwu1,luwu2,ivash,izm1,izm2,kong}.

Many critical systems have been shown to have a local scale
invariance, so that their scaling limit can be described by a
conformal f\/ield theory. Such a theory is primarily (but
non-exclusively) parametrized by the value of its central charge
$c$, which itself is related to the f\/inite-size corrections to
the critical free energy.

The calculation of the central charge based on the f\/inite-size
corrections has led to some confusion in the literature. Indeed it
is well-known since \cite{ferd} that the f\/inite-size corrections
depend on the parities of $M$ and $N$. This has prompted several
authors to claim that correspondingly the central charge was also
dependent of the parities of $M$ and $N$ \cite{tzwu}.

Here we reanalyze this question. The f\/inite-size corrections
determine univoquely the ef\/fective central charge $c_{\rm eff} =
c - 24h_{\rm min}$ \cite{itz}. To extract the central charge
itself, one needs to determine the groundstate energy $h_{\rm
min}$ of the Hamiltonian with prescribed boundary conditions. So
because of $h_{\rm min}$ the ef\/fective central charge depends on
the boundary conditions. We show, by changing variables from dimer
coverings to spanning trees (and more generally,
 to arrow conf\/igurations), that a change of parity of $M$ or $N$ has precisely
 the ef\/fect of changing the boundary conditions. The known values of $h_{\rm min}$
 and the values of $c_{\rm eff}$ as computed from the f\/inite-size corrections enable
 us to obtain a consistent value for the central charge, here equal to $c=-2$.
 Surprisingly, this is not the only consistent value of $c$ that can be used to
 describe the dimer model, since it has been shown in that a dimer conf\/iguration
 can be encoded in a height function \cite{bh}, which converges in the scaling
 limit to a Gaussian f\/ield, with $c=1$ \cite{ken}. We will discuss this peculiar situation in the Conclusion.

In what follows, we consider the dimer model on a square lattice
in the close-packed limit, and for two dif\/ferent boundary
conditions, free and periodic. In Section~2, we collect the known
results for the f\/inite-size corrections, and the way these
relate to the central charge. In Sections~3 and 4, we compute the
central charge from the f\/inite-size analysis respectively on a
strip with free boundary conditions, and on a cylinder (periodic
boundary conditions). The last section summarizes our point of
view on the description in terms of $c=-2$ versus the one based on
$c=1$, relating this issue to the more general monomer-dimer
problem. The present results have been reported in \cite{izm2},
but we take here the opportunity to give more details.

\section{Finite-size analysis}

The partition function (\ref{pf}), in the close-packed case, has
been f\/irst computed by Kasteleyn as the Pfaf\/f\/ian of a
certain matrix \cite{kast1}. Since then it has reproduced by a
variety of methods \cite{baxt}, including that based on spanning
trees \cite{temp,priez} which we will use later on.

The partition function, up to an irrelevant factor, depends on
$x/y$ only. In the inf\/inite volume limit, the dimer model on the
square lattice becomes critical but does not show a phase
transition (unlike on other lattices). The critical properties do
not depend on the values of $x/y$ (provided it is neither zero nor
inf\/inite), so we set $x=y=1$ in the following. The partition
function,
\[
Z(M,N) = \# \hbox{ dimer coverings},
\]
simply counts the number of ways the grid $\L$ (minus a corner if
$MN$ is odd) can be fully covered by dimers. The topology of $\L$
is f\/ixed by the boundary conditions: it forms a rectangle if
free boundary conditions are imposed in two directions (like in
Fig.~1a), or a cylinder if periodic boundary condition is chosen
in the horizontal direction.

We are especially interested in the free energy $F_N$ per unit of
height for an $\infty \times N$ lattice,
\[
F_N = -\lim_{M \to \infty} {1 \over M} \log{Z(M,N)}.
\]
The free energy depends on the boundary conditions, free or
periodic, and on the parity of $N$. This leads to four quantities
$F_{N,\rm even}^{\rm free}$, $F_{N,\rm odd}^{\rm free}$, $F_{N,\rm
even}^{\rm per}$ and $F_{N,\rm odd}^{\rm per}$.

The partition functions with these boundary conditions can be
expressed in terms of $Z_{\alpha,\beta}$ for $\alpha,\beta = 0,
\frac{1}{2}$ \cite{izm1}, with
\[
Z^2_{\alpha,\beta}(1,M,N) =
\prod_{n=0}^{N-1}\prod_{m=0}^{M-1}4\left[ \sin^2{\frac{\pi
(n+\alpha)}{N}} + \sin^2{\frac{\pi (m+\beta)}{M}}\right].
\]

The full expansion of $Z_{\alpha,\beta}$ for large $M$, $N$ has
been obtained in \cite{ivash}. Using these results, one can easily
obtain the dominant f\/inite-size corrections terms for the above
four free energy densities, as power series in $N$.

For an inf\/initely long strip of width $N$ with free boundary
conditions, one obtains
\begin{gather}
F_{N,\rm odd}^{\rm free} = -\frac{G}{\pi}N - \frac{G}{\pi} + {1
\over 2}\log{(1+\sqrt{2})} + \frac{\pi}{12}\frac{1}{N} + \cdots,
\label{fodd} \\
F_{N,\rm even}^{\rm free} = -\frac{G}{\pi}N - \frac{G}{\pi} + {1
\over 2}\log{(1+\sqrt{2})} - \frac{\pi}{24}\frac{1}{N} + \cdots ,
\label{feven}
\end{gather}
where $G = 0.915965$ is the Catalan constant.

The analogous results for the periodic case, i.e.~an inf\/inite
cylinder of perimeter $N$, read \cite{ferd} \be F_{N,\rm odd}^{\rm
per} = -\frac{G}{\pi} N + \frac{\pi}{12}\frac{1}{N} + \cdots,
\qquad F_{N,\rm even}^{\rm per} = -\frac{G}{\pi} N -
\frac{\pi}{6}\frac{1}{N} + \cdots. \label{p} \ee

The dominant term in these expressions yields the exponential
growth of the number of dimer coverings of a $M \times N$ grid,
namely $Z(M,N) \simeq (e^{G/\pi})^{MN} = (1.79162)^{MN}$, for both
types of boundary conditions.

The general form of the free energy per unit length of an
inf\/initely long strip of f\/inite width~$N$ at criticality is
\be F_N = f_{\rm bulk} N + 2f_{\rm surf} + \frac{A}{N} + \cdots,
\label{free} \ee where $f_{\rm bulk}$ is the free energy per bulk
site, and $f_{\rm bulk} + f_{\rm surf}$ is the free energy per
boundary site (assuming identical boundary conditions on the two
sides of the strip), and $A$ is a constant. Unlike the free energy
densities $f_{\rm bulk}$ and $f_{\rm surf}$, the constant $A$ is
universal. The value of $A$ is related to the central charge $c$
of the underlying conformal theory, and depends on the boundary
conditions in the transversal direction. The result is that $A$ is
proportional to the ef\/fective central charge $c_{\rm eff} = c -
24h_{\rm min}$ \cite{blote,aff,cardy},
\begin{gather}
A = -{\pi \over 24}\, c_{\rm eff} = \pi \left(h_{\rm
min}-\frac{c}{24}\right)
\;\; \mbox{on a strip}, \label{Astrip} \\
\noalign{\medskip} A = -{\pi \over 6}\, c_{\rm eff} = 4\pi
\left(h_{\rm min}-\frac{c}{24}\right) \;\; \mbox{on a cylinder}.
\label{Acyl}
\end{gather}
The number $h_{\rm min}$ is the smallest conformal weight in the
spectrum of the Hamiltonian with the given boundary conditions
(for the cylinder, we assumed that this operator is scalar,
$h_{\rm min} = \bar h_{\rm min}$).

\renewcommand{\thefootnote}{\arabic{footnote}}
\setcounter{footnote}{0}

The values of $A$ given above are most easily understood from the
transfer matrix formalism for calculating the lattice partition
function, even though the result seems generally more valid (i.e.\
when there is no transfer matrix). Let us f\/irst consider a
cylinder of perimeter $N$ and height~$M$, with certain boundary
conditions $|b\rangle$ and $|t\rangle$ on the bottom and top
edges\footnote{In the limit $M \to \infty$, we expect that the
specif\/ic boundary conditions we choose do not matter if the
theory is local, since the boundaries are sent of\/f to
inf\/inity. We could as well choose periodic boundary condition in
the vertical direction, closing the cylinder into a torus.}. The
transfer matrix $\cal T_{\rm cyl}$ is labelled by the degrees of
freedom living on a horizontal circle, in terms of which the
partition function is equal to $Z(M,N) = \langle b|{\cal T}_{\rm
cyl}^M|t\rangle$. In the thermodynamic limit, the vector space
spanned by the degrees of freedom living on a circle (space
coordinate) goes over to an inf\/inite Hilbert space ${\cal
H}_{\rm cyl}$, and the transfer matrix, which can be viewed as a
unit translation operator in the vertical direction (time
coordinate), can be written as ${\cal T}_{\rm cyl} = {\rm
e}^{-H_{\rm cyl}}$, in terms of a Hamiltonian $H_{\rm cyl}$. In
the large $M$ limit, the partition function $Z(M,N) = \langle
b|{\rm e}^{-MH_{\rm cyl}}|t \rangle$ will be dominated by
ground-state $E_0$ of $H_{\rm cyl}$, so that $F_N =
-\lim\limits_{M \to \infty} {1 \over M} \log{Z(M,N)} = E_0$.

The Hamiltonian is the charge associated to the time-time
component of the
 stress-energy tensor $H_{\rm cyl} = {1 \over 2\pi} \int_0^N  {\rm d}u  T_{00}$.
 If the system is critical and conformally invariant, the Fourier modes of
 the stress-energy tensor are the left and right Virasoro modes $L_n$,
 $\bar L_n$, and the Hamiltonian is simply given by the zero-th moded Virasoro
 generators as $H_{\rm cyl} = {2\pi \over N}  (L_0 + \bar L_0 - {c \over 12})$.
 This last formula assumes a normalization where the ground-state energy
 vanishes in the thermodynamic limit $N \to \infty$,
 and therefore ignores the bulk term in  (\ref{free})
 (in the periodic geometry we consider here, there is no surface and therefore
 no surface term in (\ref{free})). In addition the Hilbert space decomposes
 into representations of the left and right Virasoro algebras as ${\cal H}_{\rm cyl}
 = \oplus_{h,\bar h} \; N_{h,\bar h} \; {\cal R}_h \otimes {\cal R}_{\bar h}$,
 in which the $N_{h,\bar h}$ are integer multiplicities. The representations
 ${\cal R}_h \otimes {\cal R}_{\bar h}$ are highest weight representations,
 meaning that they are built from a highest weight state $|h\rangle \otimes |\bar h \rangle$
 by applying all Virasoro modes. All states of the representation are eigenvectors
 of $L_0 + \bar L_0$, but the state with the smallest eigenvalue is the  highest
 weight state, $(L_0 + \bar L_0 - h - \bar h)|h \rangle \otimes |\bar h \rangle = 0$.
 Putting all together, one obtains that the ground-state of $H_{\rm cyl}$ in ${\cal H}_{\rm cyl}$ is equal to
\[
E_0 = {2\pi \over N}  \left(h_{\rm min} + \bar h_{\rm min} - {c
\over 12}\right),
\]
where $h_{\rm min}$ and $\bar h_{\rm min}$ label the
representation in ${\cal H}_{\rm cyl}$ with the smallest $(L_0 +
\bar L_0)$ eigenvalue. Assuming the equality
 $h_{\rm min} = \bar h_{\rm min}$ yields the result quoted above.

For the strip, a f\/irst change is that one has to specify the
boundary conditions $a$, $b$ on the edges on the strip. The
lattice transfer matrix depends on $a$, $b$, as do the Hilbert
space ${\cal H}_{a,b}$ and the Hamitonian $H_{a,b}$ in the
thermodynamic limit, but the formula $F_N = E_0^{a,b}$ still
holds.

The second change is that there is only one copy of the Virasoro
algebra on the strip. It implies that the Hamiltonian is now equal
to $H_{a,b} = {\pi \over N} (L_0 - {c \over 24})$, and the Hilbert
space decomposition reads ${\cal H}_{a,b} = \oplus_h
 N_h^{a,b}
 {\cal R}_h$, leading directly to the value of $A$ given in (\ref{Astrip}).
 Again the normalization means that the quantum Hamiltonian ignores the bulk and surface terms in~(\ref{free}).

The f\/inite-size corrections computed above for $F^{\rm free}$
and $F^{\rm per}$ have the correct form (\ref{free}), but at
f\/irst sight look paradoxical. Comparing them with (\ref{Astrip})
and (\ref{Acyl}), we see that the ef\/fective central charge
depends on the parity of $N$. The boundary conditions do not seem
to change with the parity of $N$. This would imply that $h_{\rm
min}$ does not change either, that the central charge itself has
to change. This would be most peculiar since the central charge
also controls the bulk conformal theory.

The only way out is to accept that the boundary conditions change
with the parity of $N$, although this is not apparent in the dimer
variables. We will show in the following that the ef\/fective
central charge changes with the parity of $N$, $c_{\rm eff}=-2$
for $N$ odd and $c_{\rm eff}=1$ for $N$ even, not because the
central charge changes, but because the value of $h_{\rm min}$
changes, due to the fact that a change in the parity of $N$
ef\/fectively changes the boundary condition. On the strip, this
ef\/fect has been already noted in~\cite{brank}.

To understand this peculiarity of the dimer model, we consider a
change of variables, namely we replace dimer conf\/igurations by
arrow conf\/igurations on a sublattice. On the strip, the arrow
conf\/igurations def\/ine spanning trees, so that the dimer model
is mapped to the spanning tree model \cite{priez,brank} or,
equivalently, the Abelian sandpile model \cite{sand}. The case of
the cylinder is slightly more complicated because the arrow
conf\/igurations do not always def\/ine spanning trees. The
analysis of this case is however similar.


\section{Dimers on a strip}

Let us consider f\/irst the dimer model on the rectangular lattice
$\L$ of size $M \times N$ with free boundary conditions. Since we
are interested in the limit $M \to \infty$, the parity of $M$ will
not matter here. For simplicity, we take $M$ odd, and discuss
successively the cases $N$ odd and $N$ even.

When $M$ and $N$ are both odd, there is a well-known bijection
\cite{temp} between close-packed dimer coverings of $\L$ with one
corner removed and spanning trees on the odd-odd sublattice $\G
\subset \L$ (thus $\G$ contains the sites whose coordinates,
counted from the lower left corner, are both odd).

A dimer containing a site of $\G$, in red in Fig. 2, can be
represented as an arrow directed along the dimer from this site to
a nearest neighbour site in~$\G$. It is easy to prove that the
resulting set of arrows generates a uniquely def\/ined spanning
tree; all the arrows point to a unique root, located at the corner
which had been removed from $\L$ (see Fig.~2). Since the dimers
which do not contain a site of~$\G$ are completely f\/ixed by the
others, one has a one-to-one correspondence between dimer
coverings on $\L$ minus a corner and spanning trees on~$\G$,
rooted at the removed corner. The Kirchhof\/f theorem then
expresses the number of dimer conf\/igurations as $Z =\det
\Delta$, where $\Delta$ is the Laplacian matrix on $\G$ with
appropriate boundary conditions, see Section IV for a proof of
this result. Viewing $\G$ as a graph, one sets $\Delta_{ij} = -1$
if sites $i$ and $j$ are connected in~$\G$, and $\Delta_{ii}$ is
equal to the number of sites in~$\G$ that $i$ is connected to,
plus 1 if $i$ is connected to the root. As shown in \cite{sand},
spanning trees on $\G$, rooted at a corner, are in bijection with
the conf\/igurations of the Abelian sandpile model (ASM) on $\G$,
with closed boundary conditions on the four boundaries, the only
sink (dissipative) site being the root of the trees.

\begin{figure}[t]
\includegraphics[width = 7.1cm]{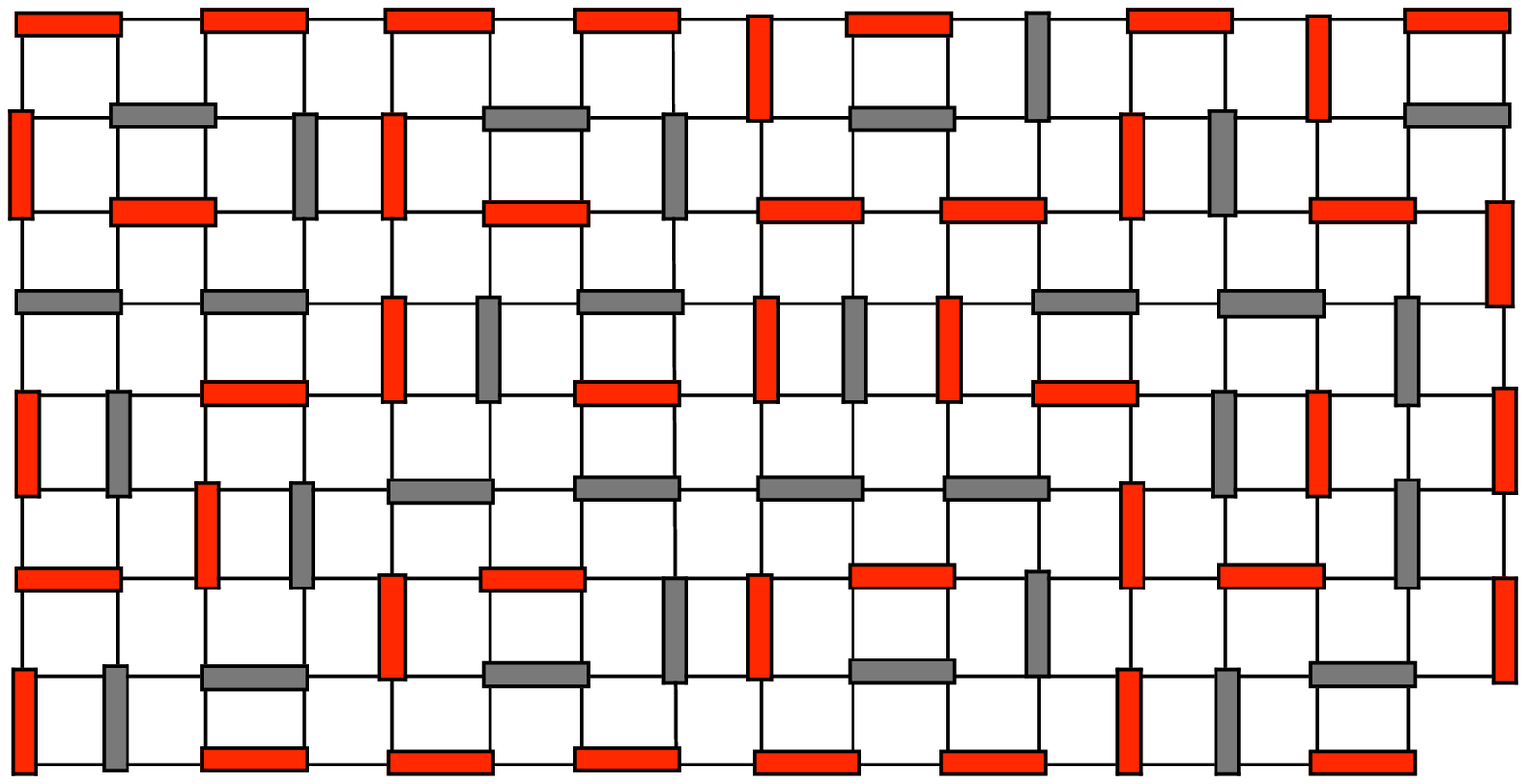}
\hfill
\includegraphics[width = 7.1cm]{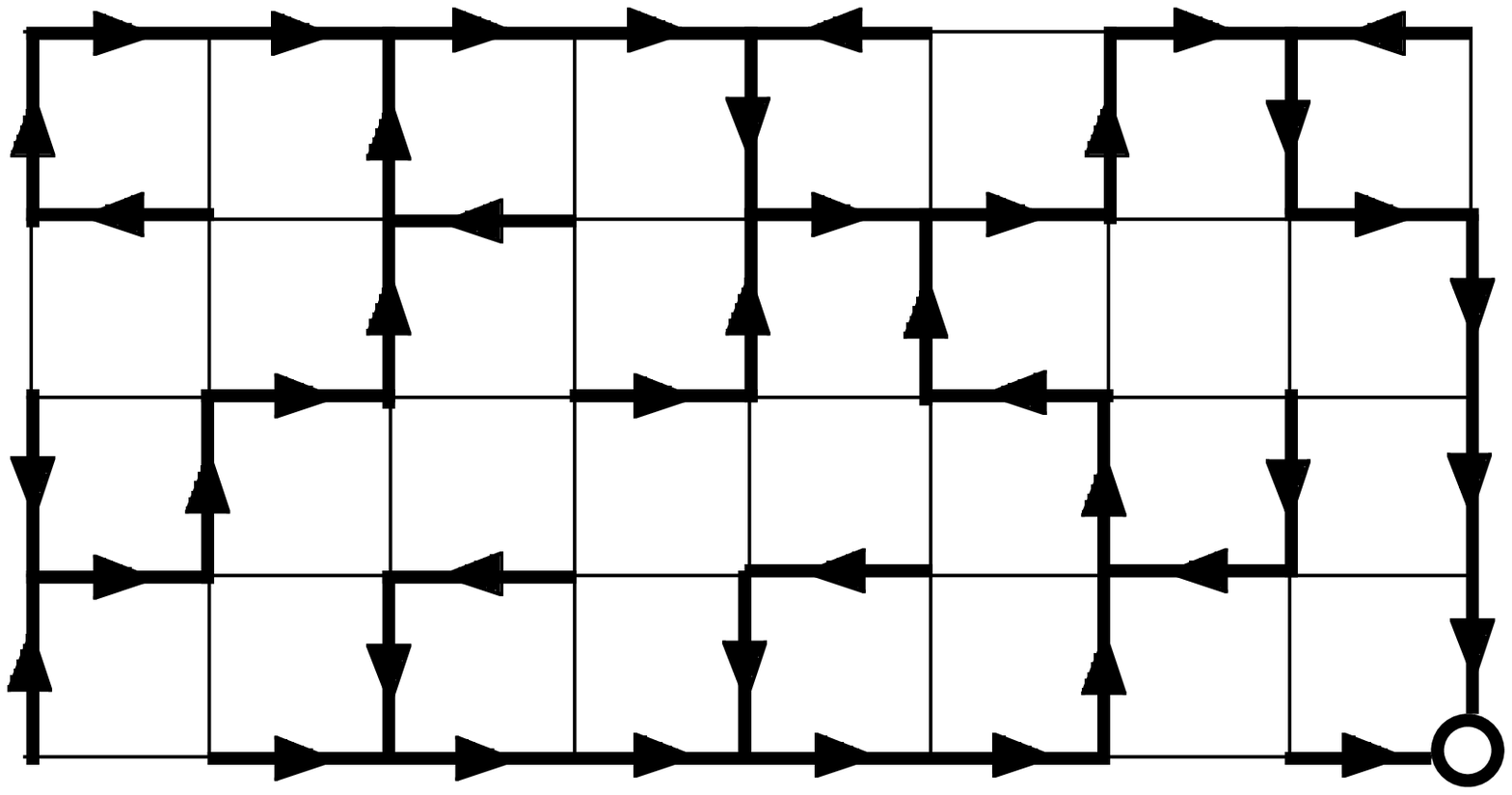}
\caption{The left f\/igure represents a dimer covering of a $9
\times 17$ lattice. The red dimers are those which touch the
odd-odd sublattice $\G$. The right f\/igure shows the
corresponding spanning tree on the $5 \times 9$ lattice $\G$, with
a unique root pictured by an open dot.}
\end{figure}

When $M \to \infty$, the lattice $\G$ becomes an inf\/initely long
strip of width $N$. The root is sent to inf\/inity so that no
boundary site (at f\/inite distance) is connected to the root. In
the ASM language, this means that the boundaries are not
dissipative and hence there is no out-f\/low of sand at the
boundaries; such boundaries conditions have been called closed
boundary conditions.

There are now very strong arguments to believe that the (scaling
limit of the) ASM on a~square lattice is described by a
logarithmic conformal f\/ield theory with a central charge $c=-2$
\cite{sand,ruelle,piru,piru1,piru2}. In particular, the spectrum
of the ASM Hamiltonian on a slice of the strip with closed
boundary conditions at the two ends has been computed in
\cite{ruelle}. There are two ground-states, the identity operator
and its logarithmic partner, both of conformal weight~0, so that
$h_{\rm min}=0$. The ef\/fective central charge in this sector is
therefore $c_{\rm eff} = -2$, and the general formula
(\ref{Astrip}) reproduces the f\/inite-size corrections
(\ref{fodd}).

When $M$ is odd and $N$ is even, dimer coverings exist without the
need to remove a corner. In this case, the above construction
leads to a set of spanning trees on the odd-odd sublattice~$\G$,
where certain arrows may point out of the lattice from the right
vertical side, see Fig.~3. Viewing this vertical boundary of $\G$
as roots for the spanning trees, we see that dimer coverings
on~$\L$ map onto spanning trees on~$\G$ which can grow from any
site of the right side. The sites on this boundary, being
connected to roots, are dissipative in the ASM language, and form
an open boundary. Thus the spanning trees map onto the ASM
conf\/igurations with one vertical open, dissipative boundary, and
the three other closed.

\begin{figure}[t]
\includegraphics[width = 7.1cm]{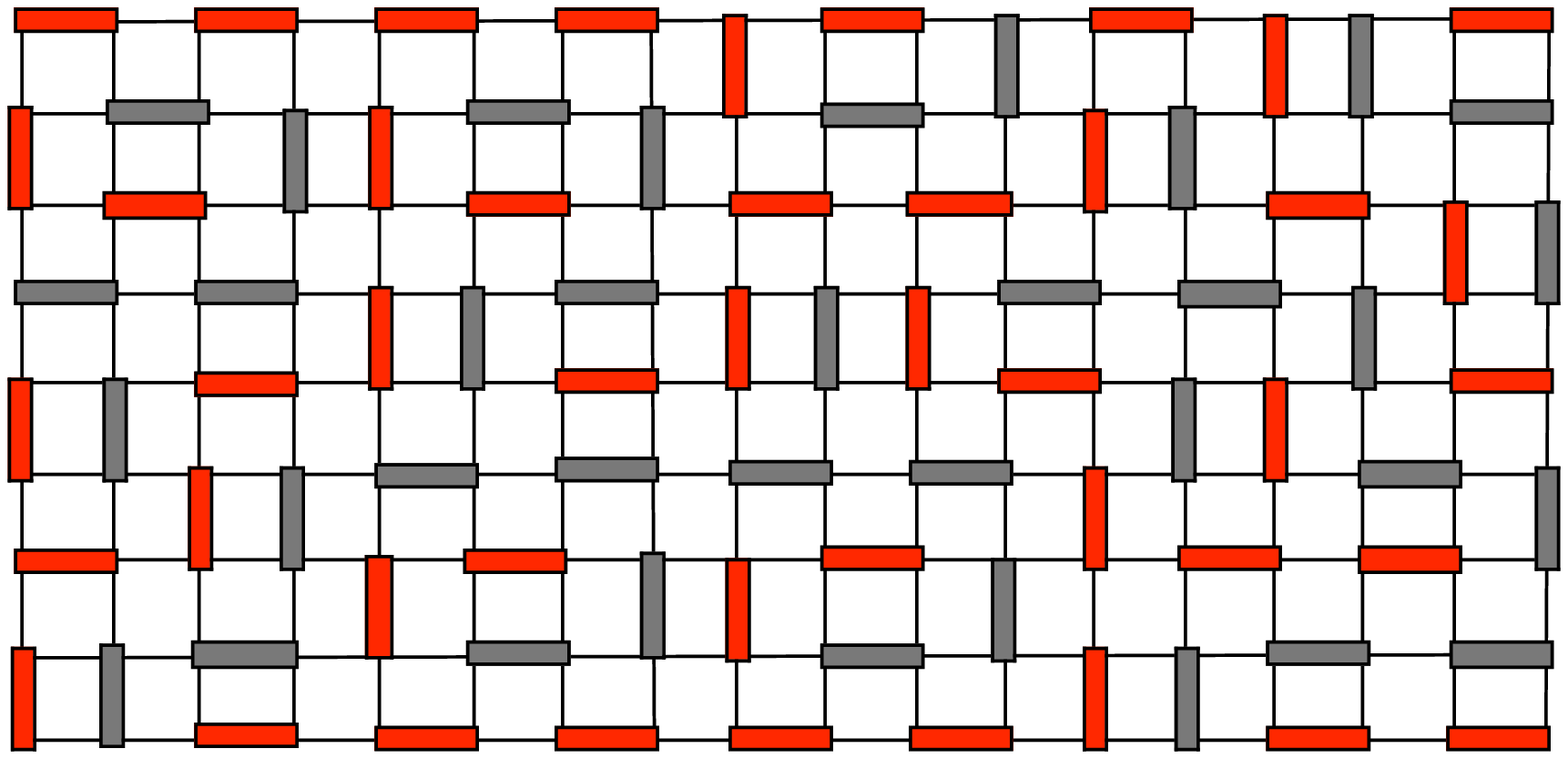}
\hfill
\includegraphics[width = 7.1cm]{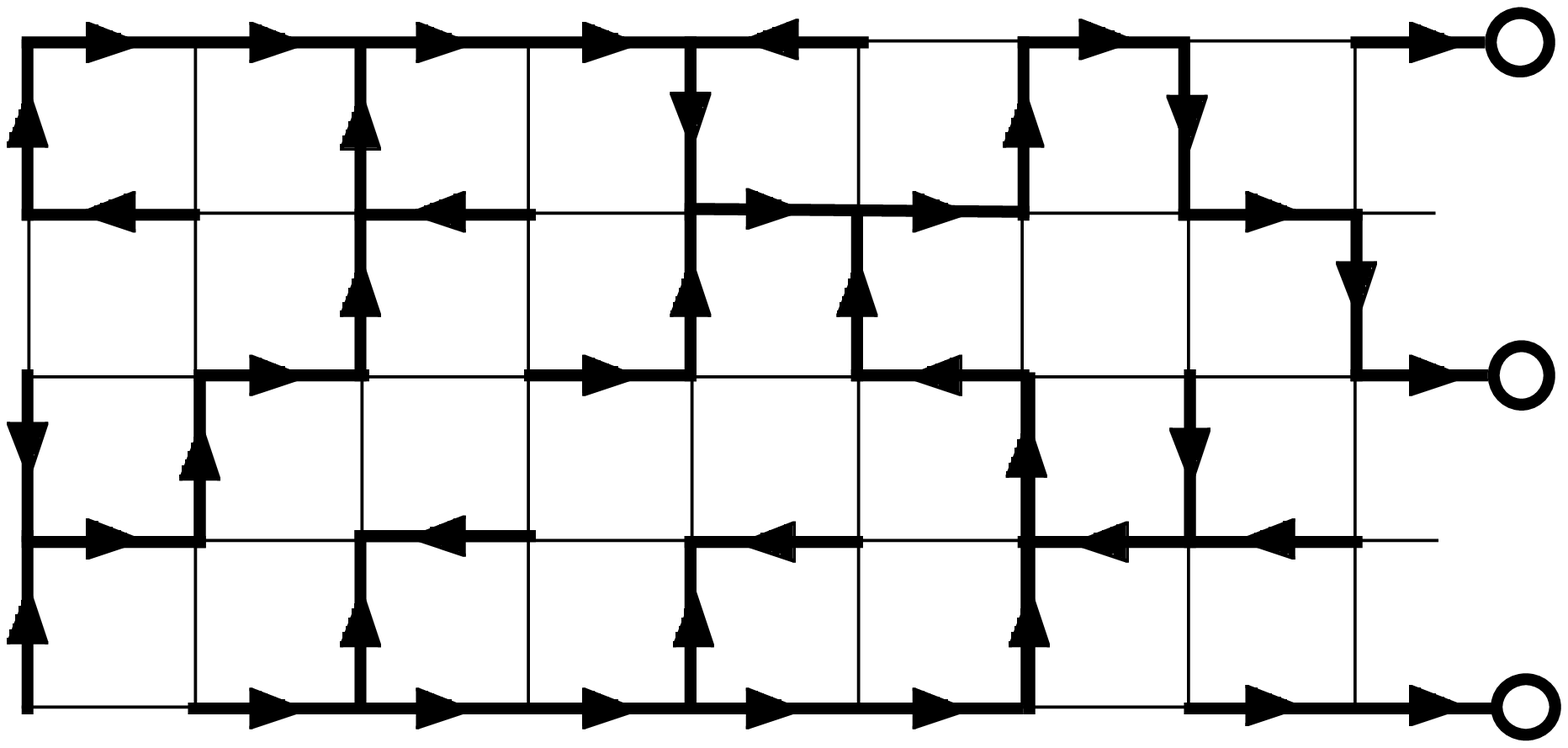}
\caption{The same dimer covering of a $9 \times 18$ lattice as in
Fig.~1 is shown on the left, where the dimers which touch the
odd-odd sublattice $\G$ are coloured in red. The right f\/igure
shows the corresponding spanning tree on the $5 \times 9$ lattice
$\G$; the sites on the right boundary are all connected to roots
(the open dots), although in this particular instance, two of
these connections have not been used by the tree.}
\end{figure}

In the limit $M \to \infty$, the lattice becomes an inf\/inite
strip with open and closed boundary conditions on the two sides.
In this case, the ground-state of the Hamiltonian with such
boundary conditions is a primary f\/ield of conformal weight
$h_{\rm min}=-1/8$ \cite{ruelle}. With $c=-2$, this yields $c_{\rm
eff} = 1$ and again the formula (\ref{Astrip}) gives the correct
result~(\ref{feven}).

Thus the leading f\/inite-size corrections for an inf\/initely
long strip of width $N$ agree with the prediction of a $c=-2$
conformal f\/ield theory, provided one realizes that changing the
parity of~$N$ genuinely changes the boundary conditions, an
ef\/fect due to the strong non-locality of the dimer
model~\cite{izm2}. The change of boundary conditions is not
apparent in the dimer model itself, but is manifest when one maps
it onto the spanning tree model or the sandpile model.

Interestingly the bijection between the dimer coverings and the
spanning trees holds if we use the even-even, even-odd or odd-even
sublattice. The boundary conditions however change, but one can
easily see what the changes are. If, instead of choosing the
odd-odd sublattice, one selects the sites whose horizontal (resp.
vertical) coordinates are even, the left and right (resp. bottom
and top) boundaries change from closed to open, and from open to
closed. So if $N$ is odd, the vertical sides become open rather
than closed. The spectrum of the corresponding Hamiltonian
changes, with a non-degenerate ground-state being the identity
operator \cite{ruelle}, so that the value $h_{\rm min}=0$ remains.
If $N$ is even, the left and right boundaries, previously closed
and open respectively, become open and closed respectively, so the
Hamiltonian remains the same, $h_{\rm min}=-1/8$. Fig.~4 shows the
spanning tree associated to the dimer covering in Fig.~3 if we
choose the even-even sublattice.

In fact the odd-odd and even-even sublattices are dual, as are the
corresponding spanning trees, see Fig.~4b. The same is true of the
odd-even and even-odd sublattices. Under this duality, the
boundary conditions open and closed are exchanged.

\begin{figure}
\includegraphics[width = 7.1cm]{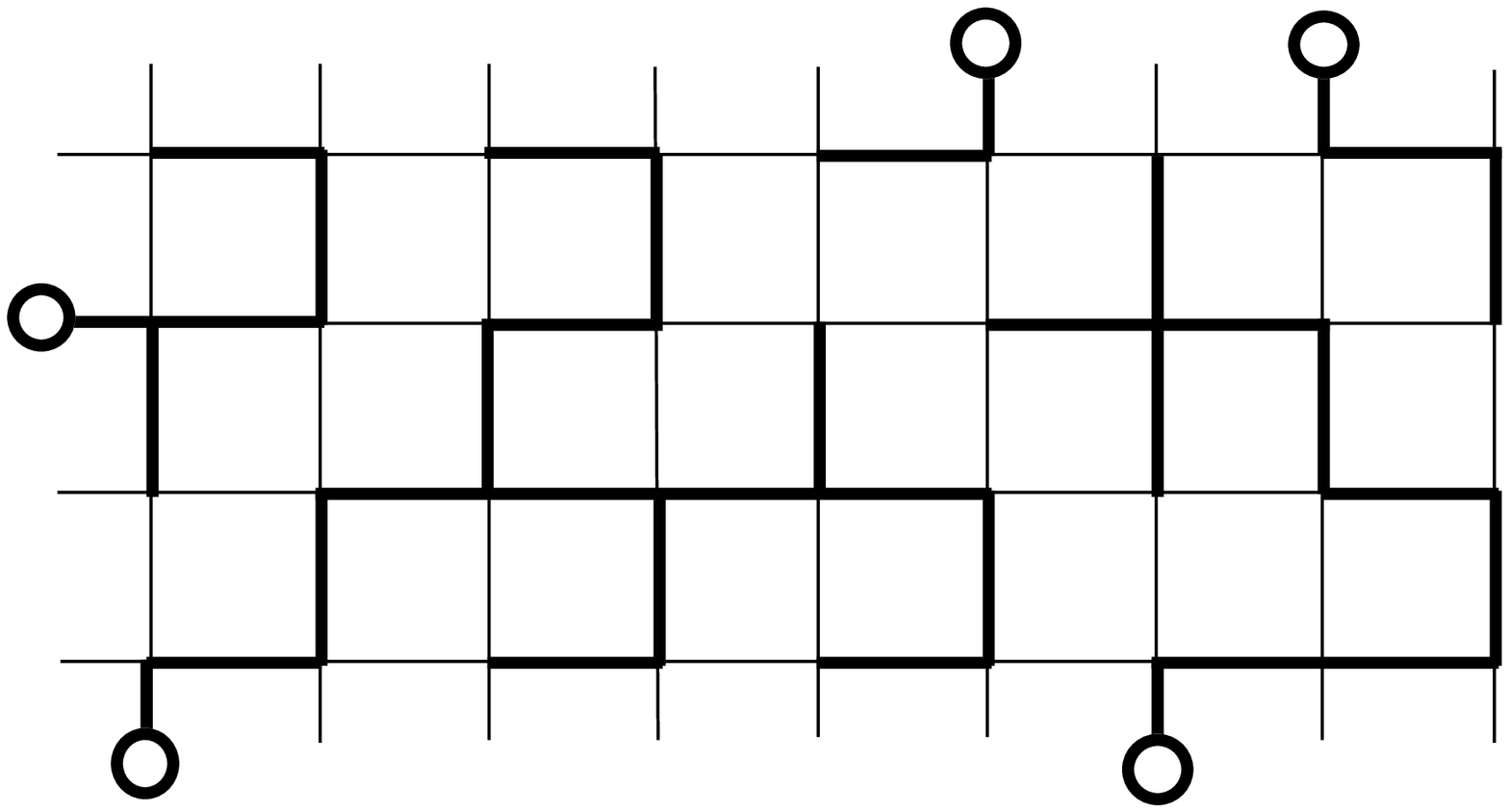}
\hfill
\includegraphics[width = 7.1cm]{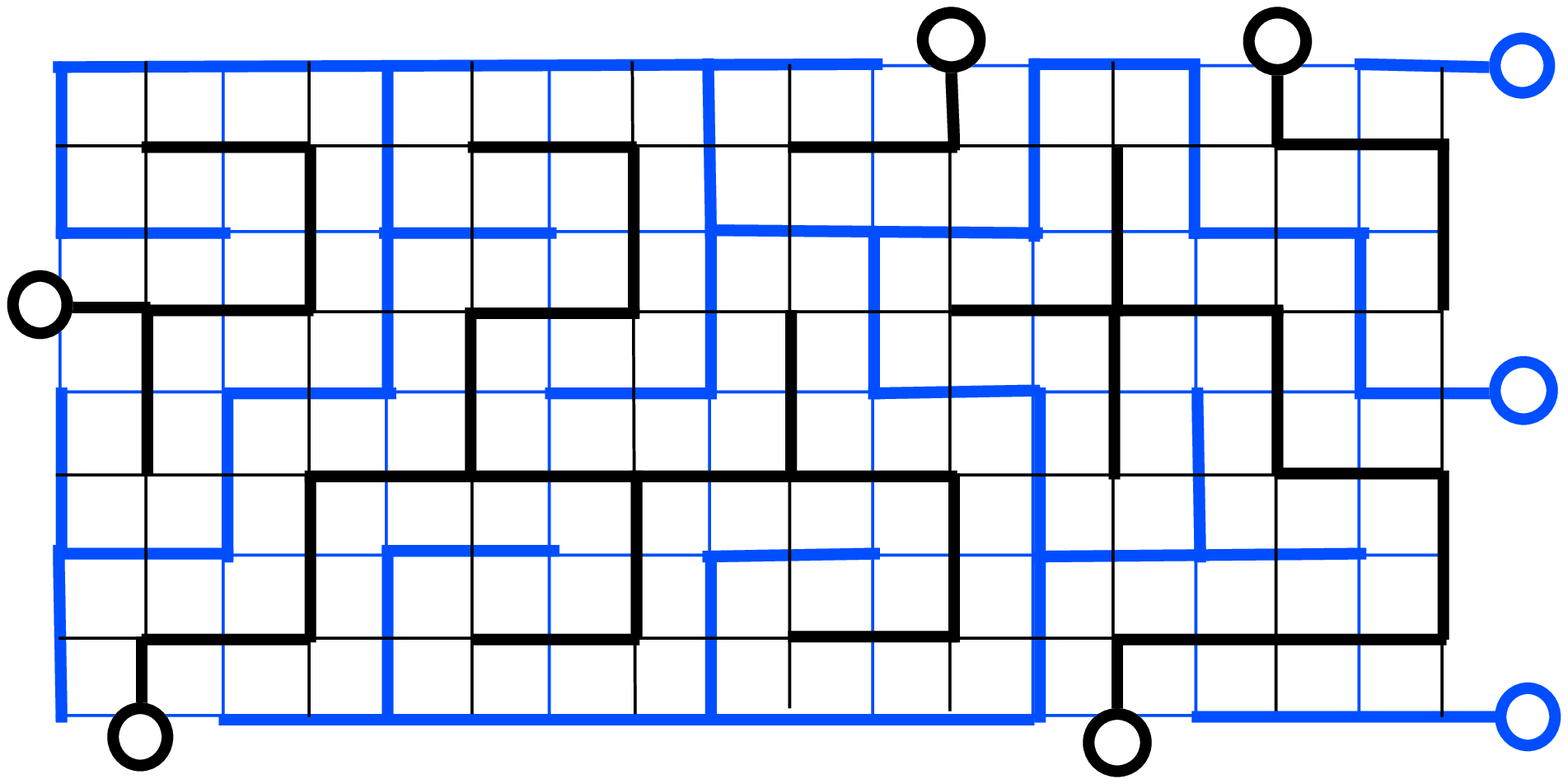}
\caption{(a)~The left f\/igure shows the spanning tree associated
to the dimer covering of Fig.~3, drawn on the even-even sublattice
rather than on the odd-odd sublattice (the arrows are no longer
shown). All boundaries are open (their sites are connected to
roots) except the right boundary, which is closed. (b)~The same
spanning tree is reproduced on the right f\/igure, where in
addition the spanning tree drawn on the odd-odd sublattice, in
blue, is superimposed. The duality of the two spanning trees is
then manifest.}
\end{figure}

\section{Dimers on a cylinder}

We consider here an $M \times N$ rectangular lattice $\L$ with
periodic boundary condition in the horizontal direction, so that
$\L$ is a cylinder of perimeter $N$ and height $M$. As before, we
will eventually take $M$ to inf\/inity, which makes its parity
irrelevant. For convenience we choose $M$ even. We discuss the
cases $N$ odd and $N$ even separately.

If $N$ is odd, we select again the odd-odd sublattice $\G$. It is
easy to see that two columns of~$\G$ will be neighbours in $\G$
and in $\L$ (connected by horizontal bonds). Therefore a dimer may
cover zero, one or two sites of $\G$. As before the dimers
covering no site of $\G$ are completely f\/ixed by the others and
play no role. For the others, we do the same construction as in
the previous section. If a dimer touches only one site of $\G$, we
draw an arrow directed along the dimer from that site to the
nearest neighbouring site of $\G$. However, for a dimer laid on
two sites of $\G$, the two arrows would point from either site to
the other, ruining the spanning tree picture. It can nevertheless
be restored in the following way.

Instead of seeing the two arrows as pointing from one site to its
neighbour, we say that they point towards roots inserted between
the neighbour sites, thus replacing the arrows \mbox{\arrowsite}
by \mbox{\arrowroot}. This in ef\/fect amounts to opening the
cylinder by removing the horizontal bonds of~$\L$ which connect
sites of $\G$, unwrapping it into a strip, and to adding columns
of roots on the left and on the right side of the strip. The new
arrow conf\/igurations def\/ine spanning trees, rooted anywhere on
the left and right boundaries. So dimer coverings on the original
cylinder are mapped to spanning trees on a strip, with open upper
and closed lower horizontal boundaries (by the arguments of the
previous section, since we chose $M$ even), and open vertical
boundaries.

However these two vertical boundaries are not independent. If a
tree grows from a site
 $i$ on the left boundary, then a tree must grow from the site $j$ on the right boundary,
 where $j$ is the site which were neighbour of $i$ in $\G$ before the opening of the cylinder.
 Conversely if no tree grows from $i$, then no tree can grow from $j$. Thus the growth patterns
 on the left and on the right boundaries must exactly match.

For f\/inite $N$, this correlation between the two boundaries will
be felt throughout the lattice. As $N$ (and $M$) increase, the
generic properties of the spanning trees will not change if the
trees are allowed to grow from slightly dif\/ferent boundary
sites. Thus if the density of roots on the left and on the right
boundaries is f\/inite, it will make no dif\/ference whether the
tree growth patterns are identical or random on each boundary.

Following this argument, when $M$ goes to inf\/inity, the lattice
becomes an inf\/inite strip with open boundary condition on either
side. As mentioned above, the ground-state of the Hamiltonian is
the identity, of weight $h_{\rm min} = 0$, leading to an
ef\/fective central charge $c_{\rm eff} = -2$. The general formula
(\ref{Astrip}) for the strip gives the correct result (\ref{p}).

This is a very unusual situation. Although the dimer model is
originally def\/ined on a cylinder, it shows the f\/inite-size
corrections expected on a strip, and must really be viewed as a
model on a strip.

For $N$ even, the problem of having a dimer occupying two
neighbouring sites of $\G$ does not arise. Hence the arrows point
from one site of $\G$ to a neighbouring site of $\G$ and they
never overlap. However the arrow conf\/igurations one obtains do
not def\/ine spanning trees because there can be sequences of
arrows looping around the cylinder. In general the arrows form
a~combination of oriented loops wrapped around the cylinder with
trees growing from the loops. The one-to-one correspondence
between the oriented loops combined with tree branches on one
side, and dimer conf\/igurations on the other side can be
established as above. The enumeration of the loop-tree
conf\/igurations requires a slight generalization of Kirchhof\/f's
theorem. In order to appreciate it, we make a little digression,
to recall the combinatorial content of the usual Kirchhof\/f
theorem (see for instance~\cite{priez}), as used in Section~3.

Def\/ine the Laplacian of an unoriented graph $\G$ as the
symmetric matrix
 $\Delta$ with entries $\Delta_{i,j}=-1$ if there is a bond connecting site
  $i$ and site $j \neq i$, and $z_i = \Delta_{i,i}$ equal to the number of sites to which
  $i$ is connected, including the root(s). Thus $\sum_j \, \Delta_{i,j}$ is the
  number of connections from $i$ to the root(s). Let us now place, at each site
  $i$ of $\G$, an arrow pointing to any one of the $z_i$ neighbours of $i$
  (possibly to the root if $i$ has a connection to it). Such a set of arrows
  def\/ines an arrow conf\/iguration on $\G$, and we want to show that the determinant
  of $\Delta$ precisely counts those arrow conf\/igurations which contain no loop.

The determinant is a sum over the permutations in $S_N$ with
$N=|\G|$ the number of sites of~$\G$, \be \det \Delta =
\sum_{\sigma \in S_N} \; \varepsilon_\sigma \:
\Delta_{1,\sigma(1)} \Delta_{2,\sigma(2)} \cdots
\Delta_{N,\sigma(N)}. \nonumber \ee Writing a permutation as a
product of cycles of lengths $\ell_m$, and using
$\varepsilon_\sigma = (-1)^{N + \#\,{\rm cycles}}$, it is not
dif\/f\/icult to see that each term in the sum carries a sign
equal to $(-1)^{\#\,{\rm proper\ cycles}}$ where by proper cycle
we mean a cycle of length strictly larger than~1. Up to this sign,
a permutation $\sigma$ with exactly $k$ proper cycles brings a
contribution equal to $\prod_i z_i$ where the product is over
those sites left invariant by the permutation. Combinatorially
this number counts the arrangements of arrows containing $k$ loops
whose locations in the graph are completely specif\/ied by the
cycle structure of $\sigma$; on the other hand, the arrows coming
out of the sites f\/ixed by $\sigma$ are free to point in any of
the allowed directions, and may therefore form themselves other
loops.

The sum over the permutations can now be reorganized as a sum over
the number $k$ of proper cycles, \be \det \Delta =
\sum_{k=0}^{[N/2]} \; (-1)^k \sum_{\sigma {\rm \ has\ } k{\rm \
proper\ cycles}} |\Delta_{1,\sigma(1)} \cdots
\Delta_{N,\sigma(N)}|. \label{altern} 
\ee

The term for $k=0$, equal to $\prod_{i \in \G} z_i$ counts the
total number of unconstrained arrow conf\/igurations. The term
$k=1$, up to the minus sign, counts the arrow conf\/igurations
which have at least one loop of a f\/ixed type, and sums over the
possible loops. It therefore overcounts the number of arrow
conf\/igurations with at least one loop by the number of arrow
conf\/igurations which have at least two loops. This is taken care
of by the $k=2$ term, which however overcounts it by a amount
equal to the number of conf\/igurations with at least three loops,
and so on. By the inclusion-exclusion principle, the alternating
sum (\ref{altern}) exactly counts the conf\/igurations of arrows
with no loop at all.

Let us now come back to the problem of counting the loop-tree
conf\/igurations of arrows, in which the only loops we allow must
wrap around the cylinder (i.e. be non-contractible). Because they
cannot intersect themselves, the loops can wrap only once around
the cylinder. For this, we modify the previous Laplacian $\Delta$
by changing from $-1$ to $+1$ the entries $\Delta_{i,j}$ for all
pairs $i,j$ such that the bond $i-j$ crosses a line going from one
boundary of the cylinder to the other boundary, which we will call
a defect line. The resulting still symmetric matrix $\Delta_A$ can
be viewed as the antiperiodic Laplacian (the lower and upper
boundaries are closed and open, or vice-versa, since we have taken
$M$ even). We show that $Z = \det{\Delta_A}$ precisely counts the
loop-tree conf\/igurations we need.

The key observation is that a contractible loop on the cylinder
crosses the defect line an even number of times, whereas a
non-contractible loop crosses it an odd number of times. As a
consequence, and with respect to the previous situation, there is
an extra minus sign for each proper cycle giving rise to a
non-contractible loop of arrows. One thus obtain
\[
\det \Delta_A = \sum_{k=0}^{[N/2]} (-1)^k \sum_{\sigma {\rm \ has\
}k{\rm \ proper\ cycles}}
 (-1)^{\#\,{\rm non-contr.}} \:|\Delta_{1,\sigma(1)} \cdots \Delta_{N,\sigma(N)}|.
\]
If we write $k=p+q$, where $p$ is the number of non-contractible
loops (NCL) and $q$ is the number of contractible loops (CL), the
summation over $k$ is replaced by two summations over $p$ and $q$,
\[
\det \Delta_A = \sum_{p \geq 0} \; \sum_{q \geq 0}  (-1)^q
 \sum_{\sigma {\rm \ has\ }p{\rm \ NCL},  q{\rm \ CL}} |\Delta_{1,\sigma(1)} \cdots \Delta_{N,\sigma(N)}|.
\]
One sees that the arrow conf\/igurations with a f\/ixed number $p$
of non-contractible loops, i.e.\ the terms with $q=0$, are all
counted positively. Moreover, for each $p$, the inner summation
over $q$ is an alternating sum which implements the
inclusion-exclusion principle and removes all contractible loops.
Therefore $\det \Delta_A$ exactly counts the arrow
conf\/igurations with no contractible loops, as claimed.

In the continuum limit, it becomes the partition function of a
free theory of antiperiodic Grassmannian f\/ields which, in turn,
gives $c=-2$ and $h_{\rm min}=-1/8$ \cite{ruelle}. This is again
consistent, since the f\/inite size correction (\ref{p}) together
with the general formula (\ref{Acyl}) for the cylinder yield
$c_{\rm eff}=1$.

\section{Conclusion}

By analyzing the f\/inite-size ef\/fects in terms of the
ef\/fective central charge $c_{\rm eff} = c - 24h_{\rm min}$, we
have shown that the non-local boundary ef\/fects in the
close-packed dimer model can be consistently accounted for by a
single conformal theory having central charge $c=-2$. We have
provided a consistent framework for understanding the dependence
of the f\/inite-size ef\/fects upon the boundary conditions.
However this should not be taken as a proof that $c$ must be equal
to $-2$. Indeed since the ef\/fective central charge merely
determines some combination of~$c$ and~$h_{\rm min}$, one cannot
obtain the values of both without some assumption about one of
them. This assumption can be a posteriori justif\/ied if the
conformal description obtained from it is fully consistent.

It turns out in this case that another consistent conformal
description exists, with $c=1$ \cite{bh,ken}, although a detailed
analysis of boundary conditions and parity dependence ef\/fects
has not been carried out in this context. Our explanation for this
curious fact is that the $c=1$ theory not only describes the
close-packed dimer model, but the general monomer-dimer model (it
has been suggested in \cite{gdj} that trimers would require
$c=2$).

On one hand, this has been illustrated for instance in \cite{fend}
where the monomer 2-point correlation function \cite{fist} has
been interpreted in terms of two uncoupled Majorana fermions. The
same interpretation can be made for the general $n$-point function
for monomers on a bounda\-ry~\cite{prru}. On the other hand, the
spanning tree description leading to the value $c=-2$ cannot
describe dimers with monomers in generic positions, for the basic
reason that they are def\/ined on a~sublattice and therefore
cannot keep track of the positions of all the monomers.

Thus the conformal theory with $c=-2$ must be viewed as a
subtheory of that with $c=1$, as it is able to describe the
degrees of freedom corresponding to dimer coverings but not
general monomer insertions. For those degrees of freedom, the two
descriptions should be equivalent.

\subsection*{Acknowledgements}

P.R.~is f\/inancially supported by the Belgian Fonds National de
la Recherche Scientif\/ique (FNRS).

\pdfbookmark[1]{References}{ref}
\LastPageEnding

\end{document}